\documentclass[aps,prd,onecolumn,12pt,tightenlines,notitlepage,groupedaddress,longbibliography]{revtex4-2}

\usepackage{amsmath,amssymb,amsfonts}
\usepackage[T1]{fontenc}
\usepackage{graphicx}
\usepackage{array}
\newcolumntype{L}{>{\centering\arraybackslash}p{3cm}}
\usepackage{adjustbox}
\usepackage{multirow}
\usepackage{hyperref}
\usepackage{booktabs}
\usepackage{braket}

\begin{document}

\title{Mirror symmetry for new physics beyond the Standard Model in $4D$ spacetime}


\author{Wanpeng Tan}
\email[]{wtan@nd.edu}
\affiliation{Department of Physics and Astronomy, Institute for Structure and Nuclear Astrophysics (ISNAP), University of Notre Dame, Notre Dame, Indiana 46556, USA}

\date{\today}

\begin{abstract}
The two discrete generators of the full Lorentz group $O(1,3)$ in $4D$ spacetime are typically chosen to be parity inversion symmetry $P$ and time reversal symmetry $T$, which are responsible for the four topologically separate components of $O(1,3)$. Under general considerations of quantum field theory (QFT) with internal degrees of freedom, mirror symmetry is a natural extension of $P$, while $CP$ symmetry resembles $T$ in spacetime. In particular, mirror symmetry is critical as it doubles the full Dirac fermion representation in QFT and essentially introduces a new sector of mirror particles. Its close connection to T-duality and Calabi-Yau mirror symmetry in string theory is clarified. Extension beyond the Standard model can then be constructed using both left- and right-handed heterotic strings guided by mirror symmetry. Many important implications such as supersymmetry, chiral anomalies, topological transitions, Higgs, neutrinos, and dark energy, are discussed.

\end{abstract}


\maketitle

\bigskip

\section{Introduction\label{intro}}

Physical laws of the nature are Lorentz invariant, as Einstein first realized in his theory of special relativity. The symmetry group resulting from this invariance is called the Lorentz group or $O(1,3)$. From a geometrical point of view, the Lorentz group is the metric-preserving holonomy group of the underlying Lorentzian (pseudo-Riemannian) manifold or our $4D$ spacetime. By definition, it includes all linear transformations $\Lambda$ that preserve the Minkowski metric $\eta$ as follows: $\Lambda^T\eta\Lambda=\eta$. For convenience, (sub)groups like orientable $SO(1,3)$ and proper orthochronous $SO^+(1,3)$, or its universal cover $SL(2,C)$, are often referred to as the Lorentz group in literature.

However, the full Lorentz group should be $O(1,3)$ which has four topologically separate components corresponding to the Klein four-group of the quotient $O(1,3)/SO^+(1,3)$. For the simple case of pure $4D$ spacetime without internal degrees of freedom (DoFs), the two discrete generators of the Lorentz group are often represented by the operators of space inversion (parity) $P$ and time reversal $T$. The four components of $O(1,3)$ can then be represented by $I$ (identity), $P$, $T$, and $PT$, respectively. Note that the two components of $P$ and $T$ do not preserve orientation of spacetime (i.e., they have a negative determinant).

In a different representation, a more interesting discrete generator would be from the orientation symmetry that governs the sign of the determinant of orthogonal transformations, which we will refer to later as the mirror symmetry $M$ for local spaces. Such an orientation symmetry comes naturally from the subgroup $O(1)=Z_2$ of $O(1,3)$ and its significance has largely been ignored in literature. It does not have any advantage over the above representation when considering pure $4D$ spacetime with no internal DoFs. On the other hand, not only are pseudo-Riemannian spacetime orientable, but any additional local spaces (e.g., fiber space of a bundle) are also naturally orientable. This makes orientation or mirror symmetry indispensable as both spacetime and local spaces should observe the full Lorentz symmetry at least formally.

In modern quantum field theory (QFT), elementary particle fields serve as internal degrees of freedom (governed by internal spaces) making $P$ and $T$ poor choices for discrete generators of the Lorentz group. For example, the conventional definition of parity operation when applied to fermions $P=\gamma^0$ does not change orientation in local spaces as the four-dimensional Dirac gamma matrices all have determinant $+1$. Neither does charge conjugation $C=i\gamma^2\gamma^0$. These and other discrete operators defined in conventional QFT seem to be for a single local orientation only. This indicates that a new type of orientation-related symmetry (i.e., mirror symmetry) should be introduced in QFT and it should introduce a new sector of particles as well.

Therefore, we can see that $CP$ and mirror symmetries are the natural choices of discrete generators of the Klein four-group in QFT. In this case, $CP$ represents the symmetry between particles and anti-particles within the same orientation sector or time reversal symmetry in the sense of regarding anti-particles as particles traveling backwards in time. Mirror symmetry $M$ as a local orientation symmetry defines two sectors of ordinary and mirror particles. In this way, $CP$ replaces time reversal symmetry $T$ but preserves orientation locally. All elementary particle fields are then divided into four sectors: ordinary particles, ordinary anti-particles, mirror particles, and mirror anti-particles.

The existence of a mirror sector of the Universe has been conjectured since Lee and Yang published their Nobel Prize-winning work on parity violation \cite{lee1956} that was later experimentally verified by Wu's group \cite{wu1957}. In the 1960s, Wigner first realized, when lecturing at a summer school, that the discrete symmetries of the Lorentz group could double the number of known particle states \cite{wigner1964} (which was also discussed further by Weinberg in his classic textbook on QFT \cite{weinberg1995}). Motivated by parity violation and later revealed $CP$ violation, three Soviet Union scientists, Kobzarev, Okun, and Pomeranchuk, proposed the concept of mirror symmetry --- it is conceivable that there exist two sectors of particles sharing the same gravity but governed by two separate gauge groups under $4D$ spacetime \cite{kobzarev1966}.

After some silence, the idea of mirror matter was revived in the 1980s mainly from interesting perspectives in astrophysics and cosmology \cite{blinnikov1983,kolb1985,hodges1993}. Later attempts to introduce ad hoc feeble interactions between the two sectors might have been too conservative \cite{berezhiani2006,cui2012,foot2014}. Most recent works \cite{tan2019,tan2019a,tan2019b,tan2019c,tan2019d,tan2020d,tan2019e,tan2020,tan2020a,tan2020b,tan2021,tan2021a,tan2023} with a new understanding of mirror symmetry, supersymmetry (SUSY), and dimensional phase transitions of spacetime could potentially solve a variety of puzzles in fundamental physics and cosmology and may indeed lead us to new physics beyond the Standard Model we have been looking for.

Here we will further discuss the deep meanings and profound consequences of mirror symmetry, in particular, in connection to string theory \cite{polchinski1998,*polchinski1998a}. The ideas presented here will strengthen the close connections and provide a clear, coherent picture to integrate all the recent works \cite{tan2019,tan2019a,tan2019b,tan2019c,tan2019d,tan2020d,tan2019e,tan2020,tan2020a,tan2020b,tan2021,tan2021a,tan2023}, forming a robust foundation for the framework of the new mirror matter theory that is readily testable in a laboratory \cite{tan2019d,tan2020d,tan2023}. Earlier mirror matter models \cite{berezhiani2006,cui2012,foot2014}, which introduced both mirror symmetry and sector-crossing interactions in an ad hoc manner, tended to focus on a single problem or limited issues in the field (mostly dark matter and neutron lifetime issues). In contrast, the new framework has the potential to consistently and quantitatively solve a much broader range of puzzles in fundamental physics and cosmology, including the origin of dark energy \cite{tan2019e}, dark matter \cite{tan2019,tan2019c}, baryon asymmetry in the early universe \cite{tan2019c}, stellar evolution and synthesis of elements \cite{tan2019a}, ultrahigh-energy cosmic rays \cite{tan2019b}, dimensional transitions of spacetime \cite{tan2020,tan2020a,tan2021} and black holes \cite{tan2020b,tan2021a}, neutron lifetime anomaly and CKM unitarity \cite{tan2019,tan2019d,tan2020d,tan2023}, and more. In particular, this article demonstrates that supersymmetric mirror models can be naturally constructed from heterotic strings of string theory, and it also presents further insights and better understanding of dark energy, supersymmetry, chiral anomalies, topological transitions, Higgs, and neutrinos.

\section{Irreducible representations of elementary particles and mirror symmetry\label{irreps}}

Irreducible representations of the full $O(1,n)$ group (see Appendix \ref{app1}) have rarely been discussed \cite{wigner1964,weinberg1995,ma2007}, and instead, representations in smaller subgroup are more often mentioned.
Considering $4D$ spacetime, all elementary particles such as matter fermions and gauge bosons should be irreducible representations of the full Lorentz group. It is well known that finite dimensional irreducible representations of group $SL(2,C)$ can be written as $(m,n)$ where $m,n=0,1/2,1 ...$. However, these are not necessarily representations in the full Lorentz group due to additional discrete symmetries. But they can induce irreducible representations of $O(1,3)$ in
the form of $(m,m)$ and $(m,n)\oplus(n,m)$ with $m\neq n$ (see Appendix \ref{app1.4}) \cite{ma2007}.

In particular, $(1/2,0)\oplus(0,1/2)$ as a Dirac spinor representation for Dirac fermions is reducible in $SL(2,C)$ but irreducible in the Lorentz group. As a matter of fact, it gives two inequivalent representations \cite{ma2007}, which correspond to two sectors of ordinary and mirror fermions under a suitable basis. Under the proposed mirror symmetry, the two representations can be combined into one set of elementary fermion particles as an irreducible fundamental representation of the full Lorentz group under QFT,
\begin{equation}\label{fullferm}
(1/2,0)\oplus(0,1/2)\oplus(1/2,0)'\oplus(0,1/2)'
\end{equation}
where each summand corresponds to ordinary fermions, ordinary anti-fermions, mirror fermions, and mirror anti-fermions, respectively. Under this basis, $CP$ transforms between particles and anti-particles while $M$ transforms between ordinary and mirror counterparts that will hereafter be distinguished by the prime symbol.

Similarly, $(1,0)\oplus(0,1)$ gives two inequivalent representations in the Lorentz group for gauge bosons \cite{ma2007}. We can again combine them into one set of elementary gauge bosons as an irreducible adjoint representation of the full Lorentz group under QFT,
\begin{equation}
(1,0)\oplus(0,1)\oplus(1,0)'\oplus(0,1)'
\end{equation}
representing gauge bosons of both the ordinary and mirror sectors.

For spacetime tensors, $(m,m)$ gives rise to four inequivalent irreducible representations in $O(1,3)$ \cite{ma2007}. For example, $(0,0)$ represents four different types of scalars: an ordinary scalar $(0,0)^+$, an ordinary pseudoscalar $(0,0)^-$, a mirror scalar $(0,0)^{+\prime}$, and a mirror pseudoscalar $(0,0)^{-\prime}$.

With respect to extended $4D$ spacetime, internal spaces exhibit two distinct orientations that can be transformed by the mirror operator. Pure $4D$ spacetime with no internal DoFs naturally incorporates $P$ and $T$ as discrete symmetry generators of $O(1,3)$. In QFT, $CP$ essentially replaces the role of $T$ due to the $CPT$ theorem that holds within a given orientation sector. However, the second discrete generator was never properly established in QFT as the awkward $P$ symmetry defined in QFT does not serve well. In the proposed new framework, mirror symmetry $M$ is indeed the missing component and recovers the orientation symmetry locally. Although related, $P$ is fundamentally different from mirror (orientation) symmetry, especially in local spaces. 

Orientation symmetry manifests in two ways: externally in the extended spacetime manifold, and intrinsically in local or internal spaces. Conventionally, parity transformations as spatial reflections are related only to external spacetime orientation. Internally, a conventional parity operation changes the spacetime chirality of a particle (e.g., a fermion), but it does not change the internal orientation of the particle due to the determinant of the Dirac gamma matrices being equal to $+1$.

Parity, with the mirror symmetry extension, can also be understood as the inversion of the spatial coordinates of both spacetime and internal spaces. For $U(1)$, $SU(2)$, and $SU(3)$ gauge spaces, the number of spatial dimensions resulting from complexification is one, two, and three, respectively, which leads to mirror parity determinants of $-1$, $+1$, and $-1$, respectively. Combining these with the determinant of $-1$ for parity in $4D$ spacetime, we find that the overall parity determinant for $U(1)$ and $SU(3)$ is $+1$, i.e., there is no change in orientation. This helps to explain why $U(1)$ and $SU(3)$ interactions conserve parity. The situation is opposite for $SU(2)$, which leads to parity violation. More on this aspect will be discussed in the next section involving string theory.

It is natural to construct a mirror operator $\mathcal{M}$ that exchanges between the two sectors in internal spaces,
\begin{equation}\label{eq:pm}
P_M: \; \psi \longleftrightarrow \psi',
\end{equation}
while serves as a chirality operator in extended spacetime, e.g.,
\begin{equation}\label{eq:gamma}
\Gamma: \; \psi_L \longrightarrow -\psi_L, \; \psi_R \longrightarrow \psi_R
\end{equation}
where without loss of generality, we assume that left-(right-)handed fermion fields are odd (even) under chirality transformation. Then we can easily obtain $\mathcal{M}$ for fermions and Higgs-like scalars due to fermion condensation as follows \cite{tan2019e},
\begin{equation}\label{eq_m}
\mathcal{M}=P_M\Gamma : \: \psi_L \rightarrow -\psi'_L, \: \psi_R \rightarrow \psi'_R, \: \phi \rightarrow -\phi'
\end{equation}
where the scalar field $\phi$ is a condensate of fermions with opposite chiralities. For consistency, the mirror operator $P_M$ on internal spaces could be imagined as a type of ``internal chirality'' transformation, such as on internal coordinates. As a result, it transforms between the two sectors by flipping the internal orientation as follows,
\begin{equation}\label{eq:pm2}
x_L\rightarrow -x_L, x_R\rightarrow x_R \Longrightarrow \psi(x_L,x_R) \leftrightarrow \psi'(x_L,x_R)
\end{equation}
where $x_L,x_R$ are not spacetime coordinates but rather internal ones. We will gain a better understanding of this mirror symmetry by comparing with results from string theory in the following sections.

\section{Mirror symmetry and string theory\label{string}}

As discussed above, discrete operators like $P$ and $CP$ defined in conventional QFT do not change the local orientation of particles because all Dirac gamma matrices have a determinant of $+1$ and their combinations always represent transformations within the same sector. It is mirror symmetry that transforms between two distinct local orientations in internal spaces. A naive way to understand mirror symmetry is to think of it as the symmetry between inward and outward orientations of internal spaces with respect to external spacetime. To understand what exactly mirror symmetry manifests in internal spaces, we find more clues in decades of developments in string theory.

Indeed, T-duality \cite{giveon1994} in string theory is closely related to the mirror symmetry that we introduced here. It was discovered that there exists a type of radius inversion symmetry, $R \longleftrightarrow 1/R$, of a circle in compactified string theory as one exchanges winding modes with momentum modes at the same time in the dual description. It is well known that this duality can also be understood as an orientation inversion in the compactified dimension, i.e., 
\begin{equation}
X=X_L(\sigma+\tau)+X_R(\sigma-\tau) \longleftrightarrow \tilde{X} = X_L(\sigma+\tau)- X_R(\sigma-\tau) 
\end{equation}
which is equivalent to the following chirality transformation,
\begin{equation}
X_L\longrightarrow X_L, \;\; X_R \longrightarrow -X_R.
\end{equation}
T-duality then describes the symmetry between two theories under the exchange of $X\leftrightarrow \tilde{X}$.

If we identify the string worldsheet $(\sigma, \tau)$ as extended spacetime, the compactified dimension as the internal space, $X$ and $\tilde{X}$ as Higgs-like ordinary\slash mirror scalar fields $\phi$ and $\phi'$ due to fermion condensation, then we can clearly see the similarities between T-duality in string theory and mirror symmetry as demonstrated in the previous section. If we change the chirality operation slightly by flipping the signs, we can immediately recover the mirror symmetry between $X$ and $\tilde{X}$,
\begin{equation}
X_L\rightarrow -X_L, \, X_R \rightarrow X_R \; \; \Longleftrightarrow \;\; X \leftrightarrow -\tilde{X}.
\end{equation}
As a matter of fact, this is exactly the type of inward--outward local orientation symmetry that we would like to identify for our new notion of mirror symmetry. We can easily see that mirror symmetry (or T-duality in this case) is nothing but an internal chirality transformation.

Instead of relating different theories via such duality symmetries as applied in string theory, we should consider it as the mirror symmetry that connects the two sectors of particles. In the $U(1)$ or $2D$ string worldsheet space, there is an intrinsic complex structure $(\sigma, \tau) \rightarrow (z,\bar{z})$. Then the complex description of the holomorphic mode $X_L(z)$ and the anti-holomorphic mode $X_R(\bar{z})$ could be related by a parity operation on the world sheet (i.e., flipping the sign of $\sigma$). T-duality tells us that the compactified string also inherits a complex structure in $X$ and $\tilde{X}$ where $X$ could be considered ``internally holomorphic'' while $\tilde{X}$ ``internally anti-holomorphic''. The true spirit of string theory may be to require complex structures in internal spaces that could be readily induced by an almost-complex spacetime.

A remarkable conclusion from the above observation is that the two ordinary\slash mirror sectors of particles are nothing but represented by holomorphic\slash anti-holomorphic modes in their corresponding internal complex space. Mirror symmetry could also be understood as some sort of complex conjugation relating two holomorphic and anti-holomorphic worlds in internal spaces.

Now that we see the close relationship between chirality and complex conjugation transformations, we can use T-duality to construct a complex structure out of a compactified string for mirror symmetry. First, we transform a pair of external spacetime coordinates $(\sigma, \tau)$ into a pair of internal string coordinates $(x_L(\sigma+\tau),x_R(\sigma-\tau))$. Then, we obtain the corresponding equivalent complex coordinates $(x=x_R+x_L,\bar{x}=x_R-x_L)$ in the string space. T-duality then ensures the following chirality-conjugation equivalence,
\begin{equation}
x_L\rightarrow -x_L, \, x_R \rightarrow x_R \; \; \Longleftrightarrow \;\; x \leftrightarrow \bar{x}
\end{equation}
which realizes a mirror parity operation for any field $\psi$ in the compactified string space,
\begin{equation}\label{eq:pm3}
P_M: \; \psi(x) \longleftrightarrow \psi'(\bar{x}).
\end{equation}
This is exactly what is desired for mirror symmetry as demonstrated in Eq. \ref{eq:pm} and Eq. \ref{eq:pm2} in the previous section.

T-duality also confirms a naive geometrical picture of mirror symmetry, i.e., the corresponding radius inversion symmetry presents an interesting view that one sector of particles regards extended spacetime as ``outside'' while the other sector considers it as ``inside''.

String theorists have also discovered another related symmetry, also called "mirror symmetry" for Calabi-Yau (CY) manifolds. To avoid confusion with our new concept of mirror symmetry, we will refer to this as CY mirror symmetry. In fact, CY mirror symmetry, as discussed below, is also closely related to our more general concept of mirror symmetry and may be a special case of it.

In their inspiring paper, Strominger, Yau, and Zaslow \cite{strominger1996a} proposed the well-known SYZ conjecture, which states that CY mirror symmetry can be obtained through T-duality. This implies that CY mirror symmetry has to be related to certain chirality or complex conjugate operations in internal spaces as well. For $6D$ Calabi-Yau or $SU(3)$ spaces, this may be the special case of mirror symmetry that relates ordinary quarks to mirror quarks.

As internal chirality transformations, mirror symmetry can be defined in various types of internal spaces that could be related to different types of particles. A generalization of an internal chiral operation could be achieved by changing the signs of all spatial coordinates in the internal space, which effectively flips between holomorphic and anti-holomorphic modes. In internal spaces with odd complex dimensions, such as $U(1)$ and $SU(3)$ spaces, there are two distinct orientations of left-moving (holomorphic) and right-moving (anti-holomorphic) modes. However, spaces with even complex dimensions (e.g., $SU(2)$) do not have this feature as both left-moving and right-moving modes must share the same internal orientation, which may be the fundamental reason why the two ordinary\slash mirror sectors share the same set of neutrinos. The extended $4D$ spacetime has an odd number of spatial dimensions and therefore exhibits two different external orientations, leading to parity violation in weak $SU(2)$ interactions.

While string theory has provided many deep insights for fundamental physics, in the following we will primarily focus on understanding the mathematical constructs in string theory and connecting these mathematical results to physical meanings in mirror matter theory. In doing so, we may need to significantly alter the original understanding in string theory and reinterpret it within the context of the new framework. Nevertheless, such connections suggest that string theory may be a very promising and powerful tool for further developing the new mirror theory.

\section{Low-dimensional supersymmetric mirror models and string theory\label{smm2}}

One of the most astonishing achievements in superstring theory is that a critical dimension of $D=10$ is required for the target spacetime to be consistent or anomaly-free. However, there exists, though not well advertised, another critical dimension of $D=2$ in superstring theory as demonstrated below. Under the BRST formalism, it can be shown that the total central charge of the Virasoro algebra in string theory must be zero. On the one hand, superspace $(X^D,\theta^D)$ contributes a central charge of $c=D+D/2=3D/2$. On the other hand, the Faddeev-Popov ghosts, with conformal weights of $(\lambda, 1-\lambda)$, contribute a canceling central charge \cite{polchinski1998,*polchinski1998a},
\begin{equation}
c_g = -2(-1)^{2\lambda}(6\lambda^2-6\lambda+1).
\end{equation}

As shown in Table \ref{tab1}, two critical dimensions ($D=2,10$) emerge under two different scenarios. Specifically, as discussed below, the critical dimension of $D=2$ with ghosts of spin 1/2 and 1 corresponds to the supersymmetric mirror models of SMM2 and SMM2b, while the other case of $D=10$ with ghosts of spin 3/2 and 2 leads to the models of SMM4 and SMM4b.

\begin{table*}
\caption{\label{tab1} Central charge contributions from Faddeev-Popov ghosts, resulting critical dimensions, and corresponding supersymmetric mirror models are listed.}
\begin{ruledtabular}
\begin{tabular}{c c c | c | c | c}
ghost types & $(\lambda, 1-\lambda)$ & $c_g$ & $c$ (all ghosts) & critical $D$ & models\\
\hline
commuting & $(1/2,1/2)$ & -1 & \multirow{2}{*}{-3} & \multirow{2}{*}{2} & \multirow{2}{*}{SMM2 / SMM2b} \\
anti-commuting & $(1,0)$ & -2 & & & \\
\hline
commuting & $(3/2,-1/2)$ & 11 & \multirow{2}{*}{-15} & \multirow{2}{*}{10} & \multirow{2}{*}{SMM4 / SMM4b} \\
anti-commuting & $(2,-1)$ & -26 & & & \\
\end{tabular}
\end{ruledtabular}
\end{table*}

In the framework of the new mirror theory, extended spacetime can undergo dimensional transitions, leading to the emergence of different sets of particles and interactions at different dimensions of the extended spacetime \cite{tan2020,tan2021}. These transitions may occur during cosmic inflation of the early Universe and during the collapse of a massive star into a black hole at late stages of stellar evolution. With this in mind, we can consider how developments in string theory can be incorporated into the new framework.

Modern quantum field theories can be formulated under the mathematical language of fiber bundle theory \cite{wu1975}. In this context, the critical dimension of the target space in string theory can be understood as the dimension of the base manifold of a fiber bundle. For the purpose of determining these critical dimensions, the $2D$ worldsheet in string theory can be seen as the minimum requirement for universal $U(1)$ or complex structures in any internal spaces. Furthermore, the extended spacetime can be understood as part of the base manifold, with other parts possibly being compactified or curled up, while gauge interactions can be obtained from compactified spaces or fiber spaces.

The critical dimensions of $D=2$ and $D=10$ are essential for constructing consistent supersymmetric mirror models. We will focus on the $D=2$ cases in this section and then discuss the $D=10$ models in more detail in the next section.

Under the scenario of dimensional evolution or inflation of spacetime, the mirror models (SMM1 and SMM1b) for the beginning of the Universe have Lagrangians consisting of a single real scalar $\varphi$ \cite{tan2020},
\begin{equation}\label{eq:l1d}
\mathcal{L}_{\text{SMM1}} = \frac{1}{2}\dot{\varphi}^2
\end{equation}
and
\begin{equation}\label{eq:l1dssb}
\mathcal{L}_{\text{SMM1b}} = \frac{1}{2}\dot{\varphi}^2 - V(\varphi^2)
\end{equation}
which, under $1D$ (space)time, are fairly trivial with the mirror or orientation symmetry being identified as its holonomy group $O(1)$ or time reversal symmetry \cite{tan2020,tan2021}. They describe $1D$ cases before supersymmetry or string theory becomes applicable.

However, the proposed supersymmetric mirror models (SMM2 and SMM2b) in $2D$ spacetime can be directly related to string theory. Starting from $2D$, the true essence of string theory, i.e., the meaning of strings, is its natural requirement of complex structures in internal spaces of QFT. The base manifold (in this case, $2D$ spacetime) of a fiber bundle provides the stage for QFT and gravity and it is also responsible for the birth of matter fermions (see Appendix \ref{app:rep}).

Meanwhile, fiber spaces and\slash or other compactified spaces provide holonomy gauge groups for representations of gauge interactions and bosons. Scalars, in particular massive ones that provide new mass scales, emerge from the condensation of fermions of opposite chiralities, which may lead to a massive world or further dimensional transitions. Self-consistent gauge and chiral supermultiplets appear naturally and become building blocks for the supersymmetric mirror models. More details can be seen in the next section or in Appendix \ref{app:rep}.

In particular, under the critical dimension of $D=2$, a gauge supermultiplet and a chiral supermultiplet as shown in Table \ref{tab:rep} emerge and can be used to build the models of SMM2 and SMM2b. Their Euclidean actions are proposed as follows (see Appendix \ref{app2}),
\begin{equation} \label{eq:s2}
\mathcal{S}^E_{\text{SMM2}} = \int d^2z\, \{\frac{1}{4}(\bar{\partial}A(z)-\partial \bar{A}(\bar{z}))^2+\lambda(z)\bar{\partial}\lambda(z) + \bar{\lambda}(\bar{z})\partial\bar{\lambda}(\bar{z})\}
\end{equation}
which contains a simple gauge supermultiplet of free and massless Majorana fermions $\lambda, \bar{\lambda}$ and $U(1)$ bosons $A,\bar{A}$ in conformal gauge, similar to a string theory model on a $2D$ worldsheet, and
\begin{equation}\label{eq:s2b}
\mathcal{S}^E_{\text{SMM2b}}
= \int d^2z \, \{\partial\phi \bar{\partial}\phi - \lambda\bar{\partial}\lambda - \bar{\lambda}\partial\bar{\lambda} -V''(\phi)\lambda\bar{\lambda} -\frac{1}{4}(V'(\phi))^2\}
\end{equation}
which is a $N=(1,1)$ supersymmetric model where $\phi = \phi_L(z) + \phi_R(\bar{z})$ is the sum of both holomorphic and anti-holomorphic scalars that could be considered as condensation states of Majorana fermions in Eq. \ref{eq:s2}. The terms in these actions involving auxiliary fields to close the supersymmetry algebra off-shell are omitted for simplicity.

Here we use the Euclidean formalism to explicitly present the complex structures of the models. The symmetric Lagrangians between holomorphic and anti-holomorphic modes exactly demonstrate the mirror symmetry in $2D$ cases. In next section, these Abelian complex structures will be extended to non-Abelian cases using Yang-Mills gauge theories and the mirror symmetry will be further complicatedly embedded.

SMM2 has been applied to well describe the microphysics of Schwarzschild black holes as $2D$ boundaries of $4D$ spacetime \cite{tan2021a}. SMM2b could be used to explain the dynamics of cosmic inflation or black-hole collapsing processes. Strings in $2D$ string theory, or equivalently complex structures in the description of free Majorana fermions with $U(1)$ gauge under the new framework , could also be the origin of string or complex structures in higher dimensions extended from further spatial inflation.

The dynamic evolution from SMM2 to SMM2b could be understood or even constructed from certain fermion condensation models, such as the Nambu-Jona-Lasinio (NJL) mechanism using four-fermion interactions \cite{nambu1961} and possibly the Sachdev-Ye-Kitaev (SYK) model via random interactions of many fermions \cite{sachdev1993,kitaev2015,*kitaev2015a,maldacena2016a}.

\section{Supersymmetric mirror models in $4D$ spacetime and string theory\label{smm4}}

\subsection{Splitting of spaces and fermion particles\label{fermion}}

For higher dimensional supersymmetric mirror models (SMM4 and SMM4b), the base manifold with critical dimension $D=10$ is broken into two parts: a $4D$ inflated spacetime and a $6D$ compactified Calabi-Yau space. The related string theory model is the so-called heterotic model \cite{gross1985}. However, instead of selecting one chiral model, two chirally symmetric copies of heterotic strings should be combined to be consistent with the requirements of mirror symmetry. That is, a $D=26$ left-moving bosonic string combined with a $D=10$ right-moving susperstring provides the ordinary sector, while a right-moving bosonic string plus a left-moving superstring of the same dimensions gives rise to the mirror sector, i.e.,
\begin{equation}
(\textbf{Heterotic String})_{\textbf{Left}} + (\textbf{Heterotic String})_{\textbf{Right}}\Rightarrow \textbf{SMM4/SMM4b}.
\end{equation}

The reason why only four dimensions can be fully extended in the base manifold could be understood in the simple $\phi^4$ renormalization group theory (RG). Under $4D$ spacetime, RG calculations show that the $\phi^4$ term is marginal and becomes irrelevant in higher dimensions. This means that only free scalar fields can exist in $D>4$ quantum field theory, making the Higgs mechanism and inflation impossible in higher dimensions. Therefore, our extended spacetime cannot exceed four dimensions for finite or renormalizable models of massive fields. The $6D$ Calabi-Yau space out of the $10D$ base manifold must therefore be compactified for consistency. As a result, all associated spaces, including the tangent/cotangent spaces and the extra $16$ dimensions of each chiral bosonic string as a compactified fiber space, must be split into two parts as the base manifold does.

The base manifold defines matter fermions with leptons living in the extended $4D$ spacetime and quarks confined in the $6D$ Calabi-Yau or quark space. The space splitting is the fundamental reason why we have two distinct sets of fermions, i.e., leptons and quarks. On the other hand, local spaces, including the compactified part of the base manifold, define gauge interactions and corresponding gauge bosons. Note that the $6D$ quark space is dual-purpose, serving both as the space where quarks live and as the holonomy providing $SU_c(3)$ strong forces.

Baez demonstrated that the Standard Model gauge group $G_{SM}$ is precisely the holonomy group of a $10D$ Calabi-Yau manifold whose tangent spaces split into orthogonal $4D$ and $6D$ subspaces, each preserved by the complex structure and parallel transport \cite{baez2005}. In other words, $G_{SM}$ must be a natural subgroup of $SU(5)$ due to the space splitting. It uses the following isomorphism that the Georgi–Glashow $SU(5)$ grand unified theory \cite{georgi1974} also relies on,
\begin{equation}
G_{SM} = S(U(2)\times U(3)) = SU(3)\times SU(2)\times U(1) / Z_6
\end{equation}
which amazingly gives the representation of one generation of fermions in the Standard Model (see details in Ref.~\cite{baez2005} and Table \ref{tab:su5}). Note that $SU(2)$ here does not explain the chiral feature of the weak $SU_L(2)$ group in the Standard Model. Note also that $SU(3)\times SU(2)\times U(1)$ itself is not a subgroup of $SU(5)$ and $Z_6$ has to be considered for $G_{SM}$ to be the ``true'' Standard Model gauge group.

\begin{table*}
\caption{\label{tab:su5} Fundamental representation of ordinary fermions for critical dimension $D=10$ with $4D/6D$ splitting exactly overlaps with one generation of fermions in the Standard Model \cite{baez2005}. Lepton and quark space singlets are labeled with $s^2=\wedge^0\mathbb{C}^2$ and $s^3=\wedge^0\mathbb{C}^3$ and the corresponding vector reps are $v^2=\wedge^1\mathbb{C}^2$ and $v^3=\wedge^1\mathbb{C}^3$. Note that $v^2=v^{2*}$ and Hodge star operator * maps to the dual or anti-particle rep. The convention for the $L/R$ labels of anti-particles is to follow their corresponding particle's even though anti-particles have opposite chirality.}
\begin{ruledtabular}
\begin{tabular}{c r l | c | c }
exterior algebra & decomposition & SM fermions & $SU(3)\times SU(2)\times U(1)$ rep & $SU(5)$ rep \\
\hline
$\wedge^0(\mathbb{C}^3\oplus\mathbb{C}^2)$ & $s^3\otimes s^2$ & $\bar{\nu}_R$ & $\mathbf{(1,1,0)}$ &   \\
\hline
$\wedge^1(\mathbb{C}^3\oplus\mathbb{C}^2)$ & $v^3\otimes s^2$ & $d^{r,g,b}_R$ & $\mathbf{(3,1,-1/3)} $ & \multirow{2}{*}{$\mathbf{5}$} \\
& $\oplus\;\; s^3\otimes v^2$ & $\bar{e}_L, \bar{\nu}_L$ & $\mathbf{(1,2,1/2)}$ & \\
\hline
$\wedge^2(\mathbb{C}^3\oplus\mathbb{C}^2)$ & $v^{3*}\otimes s^2$ & $\bar{u}^{r,g,b}_R$ & $\mathbf{(\bar{3},1,-2/3)}$ & \multirow{3}{*}{$\mathbf{10}$} \\
& $\oplus\;\; v^3\otimes v^2$ & $d^{r,g,b}_L, u^{r,g,b}_L$ & $\mathbf{(3,2,1/6)}$ & \\
& $\oplus\;\; s^3\otimes s^{2*}$ & $\bar{e}_R$ & $\mathbf{(1,\bar{1},1)}$ & \\
\hline
$\wedge^3(\mathbb{C}^3\oplus\mathbb{C}^2)$ & $v^{3}\otimes s^{2*}$ & $u^{r,g,b}_R$ & $\mathbf{(3,\bar{1},2/3)}$ & \multirow{3}{*}{$\overline{\mathbf{10}}$} \\
& $\oplus\;\; (v^3\otimes v^2)^*$ & $\bar{d}^{r,g,b}_L, \bar{u}^{r,g,b}_L$ & $\mathbf{(\bar{3},\bar{2},-1/6)}$ & \\
& $\oplus\;\; s^{3*}\otimes s^{2}$ & $e_R$ & $\mathbf{(\bar{1},1,-1)}$ & \\
\hline
$\wedge^4(\mathbb{C}^3\oplus\mathbb{C}^2)$ & $(v^3\otimes s^2)^*$ & $\bar{d}^{r,g,b}_R$ & $\mathbf{(\bar{3},\bar{1},1/3)} $ & \multirow{2}{*}{$\overline{\mathbf{5}}$} \\
& $\oplus\;\; (s^3\otimes v^2)^*$ & $e_L, \nu_L$ & $\mathbf{(\bar{1},\bar{2},-1/2)}$ & \\
\hline
$\wedge^5(\mathbb{C}^3\oplus\mathbb{C}^2)$ & $(s^3\otimes s^2)^*$ & $\nu_R$ & $\mathbf{(\bar{1},\bar{1},0)}$ &   \\
\end{tabular}
\end{ruledtabular}
\end{table*}

Baez's approach is very effective in building up representations of fermions according to the splitting of the base manifold (see Appendix \ref{app:rep}). It uses the fact that the base manifold, with almost-complex, pseudo-Riemannian, and symplectic structures, can naturally be associated with complex tangent\slash cotangent spaces. The exterior algebra of such a complex (holomorphic) cotangent space splitting $\mathbb{C}^5\rightarrow \mathbb{C}^3\oplus\mathbb{C}^2$ then gives a fundamental representation of ordinary fermions as shown in Table \ref{tab:su5}. Meanwhile, the splitting of the complex conjugate (anti-holomorphic) cotangent space gives a similar exterior algebra representation $\wedge (\Bar{\mathbb{C}}^3\oplus \mathbb{\Bar{C}}^2)$ for mirror fermions. Each ordinary fermion and its mirror counterpart fulfill exactly the irreducible fundamental representation of the full Lorentz group as shown in Eq. \ref{fullferm} of Sect. \ref{irreps}.

From the defining representation $\mathbf{5}=(d^{r,g,b}_R, \bar{e}_L, \bar{\nu}_L)$ of $SU(5)$ in Table \ref{tab:su5}, we can clearly see why quarks are defined in the $6D$ CY space while leptons are born in $4D$ spacetime. This explains why quarks have color confinement while leptons can propagate individually in spacetime. There are a total of 32 DoFs in each sector of fermions, but only 30 of them participate in gauge interactions within their sector. The two singlets $\nu_R$ and $\bar{\nu}_R$ in the ordinary sector should participate in the $SU_R(2)$ interactions of the mirror sector. Contrary to their role in the ordinary sector, $\nu_L$ and $\bar{\nu}_L$ are gauge singlets in the mirror sector. Therefore, we say that the two sectors share the same set of neutrinos. More on this aspect and neutrino's role in chiral supermultiplets will be discussed later.

\subsection{Gauge groups and anomalies\label{gauge}}

The group $SU(3)\times SU(2)\times U(1)$ obtained from the splitting of the base manifold discussed above is not exactly the desired gauge group. In the following, we will adapt Baez's approach also for fiber spaces in order to obtain the true gauge groups, in particular, the chiral weak $SU(2)$ group, in $4D$ spacetime.

For critical dimension of $D=10$, we need to take into account all holonomy groups related to various compactified spaces. First, the holonomy of the compact string worldsheet or complex structures underlying in all internal spaces requires a universal $U_Y(1)$ gauge group with ordinary hypercharge. And its complex conjugate gives the corresponding $U'_Y(1)$ group for the mirror sector. Such $U(1)$ groups could be considered as inherited from the $U(1)$ under $2D$ spacetime. CY mirror symmetry ensures that not only two sets of quarks but also two mirrored copies of $SU(3)$ holonomy exist in the $6D$ CY space. This leads to the color gauge groups of $SU_c(3)$ for ordinary quarks and $SU'_c(3)$ for mirror quarks.

Now we need to consider more contributions from fiber spaces. The two heterotic strings provide not only a matched $10D$ base space with two sets of fermions but also two $16D$ unmatched chiral spaces to be compactified. For the ordinary sector, the 16 extra dimensions of the left-moving bosonic string have to be compactified into a fiber space with $SU_L(8)$ holonomy that has to be split with respect to the base manifold via Baez's approach. Following a general case treated in Eq. \ref{eq:split}, we can split it as,
\begin{equation}
SU_L(8)\longrightarrow SU_L(6)\times SU_L(2)\times U_L(1) / Z_6
\end{equation}
where one of the subspaces has to be a $4D$ fiber space of extended spacetime giving the group $SU_L(2)$ and the CY quark space then has to take charge of the group $SU_L(6)$. We will see that the chiral nature of these fiber groups is critical for establishing the new framework.

We can immediately identify $SU_L(2)$ as the weak gauge group of the Standard Model for the ordinary sector. However, the flavor group $SU_L(6)$ and chiral $U_L(1)$ cannot be gauged due to anomalies. Under $4D$ spacetime, a group is gaugeable only if it is free from triangle anomalies. In general, chiral $SU_L(N)^3$ anomalies do not vanish for $N\geq 3$, so $SU_L(6)$ is not gaugeable. For $U_L(1)$, we need to consider possible triangle anomalies of $U_L(1)^3$ and $SU(N)^2U_L(1)$, which result in the following anomaly cancellation conditions,
\begin{equation}\label{eq:qi}
\sum_\text{left-handed i} Q_i^3 = 0, \;\; 
\sum_\text{left-handed i} Q_i=0
\end{equation}
which unfortunately cannot be satisfied at the same time as elaborated below.

Consider the fundamental representation of $SU(8)$ after the splitting,
\begin{equation}
\mathbf{8}=\mathbf{(6,1,c/6)} \oplus \mathbf{(1,2,-c/2)}
\end{equation}
where $c$ is a normalization constant. If we take $c=2$, then we can identify $U_L(1)$ charge as the difference $B-L$ between the baryon number $B$ and the lepton number $L$, since $Q_i$ is $1/3$ for quarks and $-1$ for leptons. From Eq. \ref{eq:qi}, the second condition holds for $B-L$ conservation while the first one does not. Therefore, the ungaugeable $U_L(1)$ has to break down as follows,
\begin{equation}\label{eq:ua1}
U_L(1) \xrightarrow[]{U_A(1)\text{ breaking}} U^{B-L}_V(1)
\end{equation}
where $U^{B-L}_V(1)$ is a global symmetry for $B-L$ conservation.

The flavor group $SU_L(6)$ has two ways to break down. One is for it to be further broken in the following way,
\begin{equation}\label{eq:su6}
SU_L(6) \xrightarrow[]{\text{chiral breaking}} SU_I(2)
\end{equation} 
where $SU_I(2)$ is a gaugeable isospin symmetry for quarks as all $SU(2)$ groups are automatically anomaly-free. So in this case, the complete gauge group for the ordinary sector becomes,
\begin{equation}\label{eq:g4}
G_{\text{SMM4}}=U_Y(1)\times SU_L(2)\times SU_c(3)\times SU_I(2)
\end{equation}
which provides massless gauge bosons with DoFs of $n_b=30$. This coincides with the DoFs of one generation SM fermions $n_f=30$ without counting $\nu_R$ and $\bar{\nu}_R$ as shown in Table \ref{tab:su5} where it indeed shows that $\nu_R$ and $\bar{\nu}_R$ do not participate in any ordinary gauge interactions. Thus, we obtain an ordinary gauge supermultiplet of $n_b=n_f=30$ involving one generation SM particles and a similar supermultiplet in the mirror sector as shown in Table \ref{tab:rep}.

\begin{table*}
\caption{\label{tab:rep} Supermultiplets and related supersymmetric mirror models under critical dimensions of $D=2,10$ are listed.}
\begin{ruledtabular}
\begin{tabular}{c l l c }
critical  & gauge & chiral  & models \\
dimension & supermultiplets & supermultiplets & \\
\hline
\multirow{2}{*}{$D=2$} & one set of $(\bar{\lambda},\lambda,\bar{A},A)$ & one set & SMM2  \\
& $n_b=n_f=2$, $U(1)$ gauge & $(\bar{\lambda},\lambda,\phi_L,\phi_R)$ & SMM2b\\
\hline
\multirow{7}{*}{$D=10$} & two sets of $(\psi,A)$ \& $(\psi',A')$ - one generation & two sets &  \multirow{3}{*}{SMM4}  \\
& $n_b=n_f=30$, $U_Y(1)\times SU_L(2)\times SU_c(3)\times SU_I(2)$ & $(\bar{\nu}_R,\nu_R,\phi_u,\phi_d)$ & \\
& $n'_b=n'_f=30$, $U'_Y(1)\times SU_R(2)\times SU'_c(3)\times SU'_I(2)$ & $(\bar{\nu}_L,\nu_L,\phi'_u,\phi'_d)$ &\\
\cline{2-4}
& two sets of $(\psi,A)$ \& $(\psi',A')$ - three generations & three sets for &  \multirow{4}{*}{SMM4b}  \\
& $n_b=n_f=90$, $U_Y(1)\times SU_L(2)\times SU_c(3)$ & each sector & \\
& $n'_b=n'_f=90$, $U'_Y(1)\times SU_R(2)\times SU'_c(3)$ & $(\bar{\nu}_R,\nu_R,\phi_u,\phi_d)^{1,2,3}$ &\\
& counting pNGBs from flavor $SU(6)$ breakdown & $(\bar{\nu}_L,\nu_L,\phi'_u,\phi'_d)^{1,2,3}$ &\\
\end{tabular}
\end{ruledtabular}
\end{table*}

On the other hand, $SU_L(6)$ could also be left as a completely global symmetry for quark flavors, explaining the existence of six flavors or three generations of quarks in the Standard Model. Meanwhile, the existence of three generations of leptons can be understood as follows: when we complexify (co)tangent spaces of $4D$ spacetime, there are three ways to pair the time dimension with one of three spatial dimensions, resulting in three different (co)tangent spaces that represent three generations of leptons.

Then the ordinary gauge group in this case is exactly the well known SM group,
\begin{equation}\label{eq:g4b}
G_{\text{SMM4b}}=U_Y(1)\times SU_L(2)\times SU_c(3)
\end{equation}
where $n_b(\text{SM})=27$ after spontaneous symmetry breaking gives masses to $W^{\pm}$ and $Z^0$ bosons of $SU_L(2)$. Meanwhile, the global flavor group $SU_L(6)$ for quarks breaks down as follows,
\begin{equation}\label{eq:su6break}
SU_L(6) \xrightarrow[]{\text{chiral breaking}} SU_V(2) \times U_V^{t}(1)\times U_V^{b}(1)\times U_V^{c}(1)\times U_V^{s}(1)
\end{equation}
where $SU_V(2)$ and $U_V^{t,b,c,s}(1)$ are a set of leftover global symmetries that conserve isospin, baryon number, and $t,b,c,s$ numbers of quarks. The chiral $U_A(1)$ breaking from $U_L(1)$ as in Eq.~\ref{eq:ua1} produces no pseudo-Nambu-Goldstone boson (pNGB) as it is dynamically canceled by other flavor $U_A(1)$'s of $t,b,c,s$ quarks produced in Eq. \ref{eq:su6break}. As a result, the well-known $U_A(1)$ problem is resolved without the need for the hypothetical axion. In the end, the flavor $SU_L(6)$ breaking produces pNGBs with DoFs of 63. Combined with the DoFs of gauge bosons, we have a total of $n_b=90$. This gives us a pseudo-SUSY multiplet with $n_b=n_f=90$ for three generations of SM particles. More details on this aspect can be seen in Refs. \cite{tan2019c,tan2019e}.

Similarly, the breakdown of $SU_R(8)$ from the right-handed heterotic string gives us the gauge groups in the mirror sector. The resulting UV limit gauge group is,
\begin{equation}
G'_{\text{SMM4}}=U'_Y(1)\times SU_R(2)\times SU'_c(3)\times SU'_I(2)
\end{equation}
which forms a gauge mirror supermultiplet with $n_b=n_f=30$ involving one generation of mirror fermions, and the mirror SM gauge group
\begin{equation}\label{eq:g4bm}
G'_{\text{SMM4b}}=U'_Y(1)\times SU_R(2)\times SU'_c(3)
\end{equation}
plus a global mirror flavor symmetry $SU_R(6)$ at the low energy end that gives a pseudo-SUSY multiplet of $n_b=n_f=90$ for three generations of mirror particles.

\subsection{SMM4 and SMM4b models\label{models}}

The gauge singlets of neutrinos could be combined with Higgs-like scalars to form chiral supermultiplets. The scalars can be effectively derived from condensation like $\langle\bar{\psi}_R \psi_L\rangle$ of fermions with opposite chiralities in the gauge supermultiplet.

At the critical dimension of $D=10$, there are two possible choices of chiral supermultiplets. For the case of one generation SM particles in the UV limit, for example, in the ordinary sector (the mirror sector is similar), there are two gauge singlets of $\nu_R$ and $\bar{\nu}_R$ that can be in a chiral supermultiplet. And two scalar fields, $\phi_u$ and $\phi_d$, could be related to fermion condensates of $\langle\bar{u}_R u_L\rangle$ and $\langle\bar{d}_R d_L\rangle$, thus, giving a possible chiral supermultiplet of $n_b=n_f=2$. Because the condensates contain left-handed quarks, these Higgs-like scalars must be weak $SU_L(2)$ doublets. Following a Higgs-like mechanism similar to that of the Standard Model, the $SU_L(2)$ bosons would gain three more DoFs, which would break the gauge SUSY structure if $SU_I(2)$ remains gauged (an alternative for a global $SU(2)$ will be discussed below in the other choice).

In other words, the scalar fields would not be able to undergo condensation and would therefore have to remain massless due to the constraints of supersymmetry. This means that the UV limit model SMM4 must be massless, and therefore we only need massless gauge supermultiplets for its on-shell Lagrangian,
\begin{equation}
\mathcal{L}_{\text{SMM4}}=-\frac{1}{4}G^a_{\mu\nu}G^{a\mu\nu} + i\bar{\psi}_j \gamma^{\mu}D_{\mu} {\psi}_j  -\frac{1}{4}G^{\prime a}_{\mu\nu}G^{\prime a\mu\nu} + i\bar{\psi}^{\prime}_j \gamma^{\mu}D^{\prime}_{\mu} {\psi}^{\prime}_j
\end{equation}
where for the ordinary sector, $G^a_{\mu\nu}$ ($a=1,2,...,15$) is the gauge field strength tensor and the gauge covariant derivative $D_{\mu} = \partial_{\mu} - i g T^a A^a_{\mu}$ depends on gauge symmetry generators $T^a$ and gauge bosons $A^a_{\mu}$ as given in Eq. \ref{eq:g4}. The gauge coupling constant $g$ may be unified to be one initially at the UV limit, but it can evolve differently for different subgroups at lower energies. The 15 massless Dirac fermion fields $\psi_j$ represent exactly one generation of quarks and leptons (excluding $\nu_R$ and $\bar{\nu}_R$) as shown in Table \ref{tab:su5}.

The reason why only one generation of particles is present in the UV limit can be understood in the following way. The UV limit means that spacetime is still in the early stages of dimensional transition from $2D$ to $4D$, which means that the two new spatial dimensions have not yet been fully inflated. Therefore, there is only one way to complexify the (co)tangent spaces of this incomplete $4D$ spacetime by pairing the two original space and time dimensions together into one complex dimension. This single type of complexified (co)tangent spaces then represents one generation of leptons. Meanwhile, $SU(6)$ breaking as in Eq. \ref{eq:su6} is not global yet, so quarks also have one generation.

The mirror sector has similar terms in the Lagrangian with the following mirror transformation applied,
\begin{equation}\label{eq_m1}
\mathcal{M}: \: \psi_L \rightarrow -\psi'_L, \: \psi_R \rightarrow \psi'_R, \: A_{\mu} \rightarrow A'_{\mu}.
\end{equation}
As discussed in Sect. \ref{string}, $SU(2)$ does not preserve two different internal orientations. Therefore, neutrinos are shared between the two sectors, leading to the following neutrino degeneracy relations, 
\begin{equation}\label{eq:degen}
\nu_L = -\nu'_L \text{ and } \nu_R = \nu'_R.
\end{equation}
where neutrinos have to be neutral because $U(1)$ does preserve two internal orientations under the mirror symmetry.

The other choice of chiral supermultiplets arises due to the emergence of three generations of particles and the breakdown of the global flavor $SU(6)$ group. At lower energies, six flavors of quarks can be condensed into six Higgs-like scalars, $H_{u,d,c,s,t,b}$. Again, these scalars must participate in weak $SU_L(2)$ interactions as a doublet $H=(\phi^+,\phi)$ since each condensate must contain one left-handed quark. After undergoing spontaneous symmetry breaking and giving masses to other particles, these scalars will have a single DoF left for each and become simple real scalars, $\phi_{u,d,c,s,t,b}$, with respect to the new vacuum configuration. Along with three families of neutrino singlets, $\nu_R^{1,2,3}$ and $\bar{\nu}_R^{1,2,3}$, they form three chiral supermultiplets for the ordinary sector as shown in Table \ref{tab:rep}. The mirror sector has similar supermultiplets.

Then we can obtain the low energy model SMM4b as an extension to the Standard Model,
\begin{equation}\label{eq:l4b}
\mathcal{L}_{\text{SMM4b}} = \mathcal{L}_{\text{SM}}(A_{\mu},\psi_L,\psi_R,H) + \mathcal{L}'_{\text{SM}}(A'_{\mu},-\psi'_L,\psi'_R,-H')
\end{equation}
which includes pseudo-SUSY multiplets of gauge bosons $A_{\mu}$ from Eq. \ref{eq:g4b}, $A'_{\mu}$ from Eq. \ref{eq:g4bm}, and Dirac fermions $\psi$ and $\psi'$ of three generations. The model also includes three sets of chiral supermultiplets involving six scalars ($H$,$H'$) for each sector. The mirror symmetry observed in this model can be demonstrated by,
\begin{equation}\label{eq_m2}
\mathcal{M}: \: \psi_L \rightarrow -\psi'_L, \: \psi_R \rightarrow \psi'_R, \: A_{\mu} \rightarrow A'_{\mu}, \: H \rightarrow -H'.
\end{equation}

The Lagrangian for each sector in the SMM4b model is largely identical to that of the Standard Model, with two exceptions. The Higgs mechanism is the same, but there are six Higgs scalars in each sector of SMM4b. Early calculations showed that six Higgs particles can indeed account for the unification of running coupling constants toward the UV limit \cite{amaldi1991}. The other exception is that the Yukawa mass terms of the Dirac neutrinos shared between the two ordinary and mirror sectors in SMM4b can be obtained as follows \cite{tan2019e},
\begin{equation}\label{eq:numass}
-y(\bar{\nu}_L \nu_R \phi + \bar{\nu}'_L \nu'_R \phi' + h.c.) = -y(\bar{\nu}_L \nu_R (\phi-\phi') + h.c.)
\end{equation}
which take into account the neutrino degeneracy conditions in Eq. \ref{eq:degen}. The masses of the neutrinos are then determined by the ordinary-mirror mass splitting scale of $\langle \phi-\phi' \rangle \sim v-v'=\delta v$ with a fairly well constrained relative scale of $\delta v /v =10^{-15} \text{--} 10^{-14}$ \cite{tan2019}, which agrees very well with current experimental constraints on neutrino masses \cite{tan2019e}.

To summarize, supersymmetric mirror models in $4D$ spacetime naturally extend the Standard Model of three generations of massive particles to the UV limit of one generation of massless particles. Again, the dynamic picture from SMM4 to SMM4b can be better understood with fermion condensation mechanisms such as the NJL model \cite{nambu1961} and the SYK model \cite{sachdev1993,kitaev2015,*kitaev2015a,maldacena2016a}. The various mass scales in the Standard Model are likely due to staged quark condensation in SMM4b that gradually breaks all of its axial symmetries of the global flavor $SU_L(6)$ and $U_L(1)$ in the ordinary sector. The only leftover symmetries are approximate vector symmetries in the conservation of top, bottom, charm, strange, baryon, and $B-L$ numbers, and isospin of up and down quarks.

\subsection{Dark energy and Higgs mechanism\label{dark}}

With mirror symmetry, we understand that the ordinary sector is left-moving or holomorphic while the mirror sector is right-moving or anti-holomorphic. There are no gauge interactions connecting the two sectors, meaning that any field or physical observable from the two sectors must be harmonic, i.e., the sum of a holomorphic function and an anti-holomorphic function. This means that the ordinary gauge vacuum $\ket{0}$ and the mirror gauge vacuum $\ket{\bar{0}}$ have to satisfy the following conditions,
\begin{equation}
\chi(z) \ket{\bar{0}} = 0, \;\; \chi(\bar{z}) \ket{0} = 0
\end{equation}
for any ordinary (holomorphic) field $\chi(z)$ and mirror (anti-holomorphic) field $\chi(\bar{z})$.

When we take the vacuum expectation value of any operator or correlation functions, these two different vacuum states naturally annihilate any possible product mixing of ordinary and mirror fields, effectively breaking the SMM4b Lagrangian into two completely separate parts as shown in Eq. \ref{eq:l4b}. This holds even for Yukawa mass and Higgs potential terms.

The Higgs mechanism can be further clarified using the NJL four-fermion interaction model \cite{nambu1961}. Considering condensation from ordinary\slash mirror quarks only, all viable four-fermion interaction terms (plus their Hermitian conjugates) are like \cite{tan2019e},
\begin{equation}
\bar{q}_R q_L\bar{\psi}_L \psi_R, \;\; \bar{q}_R q_L\bar{\psi}'_L \psi'_R, \;\;
\bar{q}'_R q'_L\bar{\psi}_L \psi_R, \;\; \bar{q}'_R q'_L\bar{\psi}'_L \psi'_R
\end{equation}
where the ordinary\slash mirror Higgs fields $\phi$ and $\phi'$ are made of quark condensates such as $\braket{0|\bar{q}_R q_L|0}$ and $\braket{\bar{0}|\bar{q}'_R q'_L|\bar{0}}$. The effective total Higgs field for both sectors is then just a simple sum,
\begin{equation}
\Phi = \phi + \phi'
\end{equation}
which is obviously harmonic and
\begin{equation}
\braket{0|\Phi|0} = \braket{\phi}, \;\; \braket{\bar{0}|\Phi|\bar{0}} = \braket{\phi'}.
\end{equation}
which would cleanly separate the Yukawa mass and Higgs potential terms for the two sectors as expected.

Now we can examine the interesting quartic Higgs term that should contribute to the vacuum energy as,
\begin{equation}
\rho_{\text{vac}} \sim \lambda \braket{\Phi}^4 = \lambda \braket{\phi+\phi'}^4
\end{equation}
which, under any of the above gauge vacua, will simply reduce to the expectation value of the normal Higgs term within the corresponding sector. However, the gravitational vacuum for general relativity is probably better defined as a split-complex or hyperbolic structure (see Appendix \ref{app1.2}) of the two gauge vacua,
\begin{equation}
\ket{g} = \ket{0} + j\ket{\bar{0}},\;\; \bra{g} = \bra{0} - j\bra{\bar{0}}
\end{equation}
where $j^2=+1$. If we evaluate the quartic term under the gravitational vacuum, we obtain
\begin{equation}
\rho_{\text{vac}} \sim \lambda \braket{g|\Phi|g}^4 = \lambda (\braket{0|\phi|0}-\braket{\bar{0}|\phi'|\bar{0}})^4=\lambda (v-v')^4 \sim (10^{-3}\, \text{eV})^4
\end{equation}
which is amazingly consistent with observed dark energy density \cite{tan2019e}. In comparison with Eq. \ref{eq:numass}, it is clear that the dark energy scale and neutrino masses share the same origin from spontaneous mirror symmetry breaking of the two sectors. Dark energy is just the residual effect of spontaneous mirror symmetry breaking and so are neutrino masses. In the UV limit, there is no dark energy as quark condensation has not occurred yet.

\subsection{Chiral anomalies and topological transitions\label{topo}}

The breaking of chiral symmetries in SMM4b at lower energies leads to the existence of various chiral anomalies in the Standard Model of QFT. These anomalies are responsible for topological processes such as the ``sphaleron'' transition \cite{klinkhamer1984} which involves nine quarks and three leptons from each generation, violates $B$ and $L$ numbers by three, but conserves $B-L$. At much lower energies, there are also ``quarkiton'' transitions \cite{tan2019c}, which involve three heavy quarks and three leptons within the same generation, violate $B$ and $L$ numbers by one, and also conserve $B-L$.

More specifically, ``sphaleron'' transitions, which are associated with the $SU_L(2)^2U(1)$ anomaly due to $U_L(1)$ or $U^{B-L}_A(1)$ breaking as shown in Eq. \ref{eq:ua1}, occur at very high temperatures above or around the electroweak phase transition. ``Quarkiton'' transitions, which are associated with more chiral anomalies due to $U_A^{t,b,c,s}(1)$ breaking as shown in Eq. \ref{eq:su6break}, occur at lower temperatures, but still above or around the QCD phase transition.

Similar to these topological transitions, the leftover isospin symmetry at much lower energies, which could be gauged with $\rho$ mesons as its gauge bosons, could give rise to the $SU_V(2)^2U^B_V(1)$ anomaly, and consequently generate another type of transitions that violate baryon number $B$ by one. Considering the isospin symmetry of neutrons and protons, we do not observe any more phase transition or condensation of these baryons, so such topological transitions could in principle occur even at room temperature. But where do such transitions go?

Fortunately, the mirror sector has a similar $SU'_V(2)^2U^{B\prime}_V(1)$ anomaly. Together, these anomalies may provide a new mechanism of topological transitions that conserves $B+B'$. The most intriguing example of such a transition is the oscillation between ordinary and mirror neutrons ($n-n'$ oscillations) with $\Delta B = -\Delta B'=1$, which could explain the well-known neutron lifetime anomaly \cite{tan2019}. These oscillations could be the easiest topological way to connect the two sectors, even at room temperature. Note that $SU_V(2)$ is a symmetry for $u,d$ quarks or protons and neutrons only, and there is no such symmetry for leptons. Therefore, the $SU_V(2)^2U(1)$ anomaly does not conserve $B-L$.

Topological transitions like $n-n'$ oscillations, which result from this anomaly, conserve $B+B'$ but break $B-L$ and $B'-L'$. In contrast, sphaleron and quarkiton transitions conserve $B-L$ and $B'-L'$ but break $B+B'$. Therefore, this is consistent with the general claim that global symmetries cannot be perfectly conserved under the general principles of quantum gravity \cite{harlow2019}.

\subsection{Further insights and speculations\label{further}}

One striking feature of the gauge groups introduced above is that they exhibit a built-in supersymmetry between matter fermions and gauge bosons within each sector, because both fermion and boson degrees of freedom happen to be the same. This is completely different from conventional studies of supersymmetry, which often introduce a new set of SUSY particles. This new approach is similar to the quasi-SUSY principle that Nambu proposed, based on the observation of a matching of DoFs between fermions and bosons in many models \cite{nambu1988,nambu1988a}. It is most likely not supersymmetry but mirror symmetry that doubles the known particles in the Standard Model.

One of the first questions we may ask is where a particle's degrees of freedom originate. First of all, non-zero spin of a particle emerges from the birth of SUSY. That is why in $1D$ only a simple real scalar exists (see Eqs. \ref{eq:l1d}-\ref{eq:l1dssb}), and there are no other DoFs. That is why fermions and gauge bosons naturally appear in $2D$ spacetime simultaneously with SUSY. And their chiralities emerge with spins as well. More DoFs of a particle field, such as $CP$ and mirror dualities, and different types of leptons and quarks, start to appear in higher dimensional spacetime. For example, a compact group of $SU(2)\times SU(2)\times Z_2 \times Z_2$ derived from $4D$ spacetime, as elaborated in Appendix \ref{app1.4}, gives full degrees of freedom for a Dirac fermion including $CP$ and mirror symmetries. Splitting of the cotangent spaces results in two types of fermions: leptons and quarks. The part along $4D$ spacetime produces $SU(2)$ doublets such as $(\nu,e)$ for leptons and $(u,d)$ for quarks. The other part along the CY quark space gives the color DoFs for quarks. Finally, the global flavor $SU(6)$ group and fully inflated four dimensions of spacetime ensure the existence of three generations of fermions.

The growth of each dimension in the base manifold is related to the inflation of a massive scalar due to the condensation of fermions living in the existing base space \cite{tan2020,tan2021}. For example, two additional spatial dimensions are extended into $4D$ spacetime due to the inflation of two scalars, resulting from Majorana fermion condensation in $2D$ spacetime. The unitarity of such inflation processes could be related to Maldacena's AdS/CFT correspondence \cite{maldacena1998} or some more general holographic principle. However, $4D$ spacetime seems to be the end of such spatial growth. The $6D$ CY quark space could not be extended due to renormalization constraints, and the related six Higgs particles from quark condensation therefore could not lead to further inflation.

Many efforts have been devoted to searching a quantum theory of gravity. Perhaps we should accept quantum\slash gravity duality as a possibility --- classical gravity lives in the extended spacetime while quantum physics originates from compact spaces of the spacetime bundle. As shown in Table \ref{tab1}, the graviton (spin=2) and the gravitino (spin=3/2) are ghosts particles and do not manifest as physical ones under $4D$ spacetime.

Since particles are born through spacetime dimensional transitions, we can further relate them to geometrical constructs. For example, the real scalar in $1D$ as shown in Eqs. \ref{eq:l1d}-\ref{eq:l1dssb} is probably the only point particle. Particles as extended objects are likely to appear first in $2D$ spacetime as strings or 1-branes. We can imagine that, in $4D$ spacetime, quarks might be some 3-branes (e.g., D3-branes) confined in the $6D$ CY space, while leptons might be 2-branes living in $4D$ spacetime. Furthermore, 2-branes could be connected by open 1-strings ($U(1)$ gauge) and open 2-strings ($SU(2)$ gauge), while 3-branes could be tied with both types of strings in addition to open 3-strings ($SU(3)$ gauge). An interesting string-net condensation theory could also shed light on such studies \cite{levin2005}.

\section{Conclusions and Outlook\label{outlook}}

Mirror symmetry as an indispensable feature in the full Lorentz group and beyond is demonstrated. T-duality and Calabi-Yau mirror symmetry in string theory could be considered as concrete realizations of the more general concept of mirror symmetry. Many other connections between supersymmetric mirror models and string theory are discussed. In particular, the use of both chiral heterotic strings makes the new models of SMM4 and SMM4b much more attractive as an extension to the Standard Model until the UV limit. A very impressive list of long-standing puzzles may potentially be understood within the new framework.

In addition to providing a solid foundation for coherently connecting various relevant mirror matter studies \cite{tan2019,tan2019a,tan2019b,tan2019c,tan2019d,tan2020d,tan2019e,tan2020,tan2020a,tan2020b,tan2021,tan2021a,tan2023} under the same theoretical framework, this paper also further investigates and clarifies a number of outstanding issues in fundamental physics and cosmology. Specifically, it elucidates how both dark energy and neutrino mass can be understood as the residual effect of spontaneous mirror symmetry breaking. Furthermore, it offers a natural explanation for color confinement, three generations of elementary fermions, and the existence of gauge groups. The new theory incorporates supersymmetry as a built-in feature in both the ordinary and mirror sectors, relating matter fermions and gauge bosons. It introduces a Higgs mechanism involving six scalars due to staged quark condensation, which is consistent with the unification of the running coupling constants towards the UV limit. The underlying physics is also presented to explain chiral anomalies and topological transitions, which in turn lead to ordinary-mirror oscillations of neutral hadrons as the basis for solving many other puzzles.

String theory is clearly a very powerful mathematical tool for further developing the new mirror matter theory. Other proposed quantum gravity theories and fermion condensation models may also be very useful in understanding the dynamics of dimensional phase transitions of spacetime. Most intriguingly, various feasible experiments \cite{tan2023} are proposed to test concrete unique predictions of the new theory, including measurement of neutron lifetime anomalies in narrow magnetic traps or under super-strong magnetic fields, and detection of unexpectedly large branching fractions of invisible decays of long-lived neutral hadrons \cite{tan2019d,tan2020d}. All these tests, to stress again, are ready to be conducted with the current technology. We may be just starting to unveil the beauty of the mirrored universe.

\begin{acknowledgments}
This work is supported in part by the faculty research support program at the University of Notre Dame.
\end{acknowledgments}

\appendix
\section{Lorentz Invariance\label{app1}}
Lorentz invariance of an n-dimensional Lorentzian manifold (e.g., spacetime) is described by the holonomy group of the manifold, i.e., the metric-preserving group $O(1,n-1)=\{\Lambda | \Lambda^T\eta\Lambda = \eta\}$ where $\eta$ is the Minkowski metric. The subgroup $O(1)$ of $O(1,n-1)$ corresponds to the orientation symmetry that is closely related to the mirror symmetry being discussed. We will elaborate on the cases of $n=1,2,4$ below, as they are the most interesting.
\subsection{$1D$ (space)time\label{app1.1}}
In the case of a $1D$ time configuration, Lorentz invariance is simple and the corresponding Lorentz group is $O(1)$, which is equivalent to the cyclic group $Z_2=\{1,-1\}$. In this case, mirror symmetry as an orientation symmetry of the manifold (i.e., flipping the sign of the determinant) coincides with its full Lorentz group, which is also equivalent to the time reversal symmetry. 
\subsection{$2D$ spacetime\label{app1.2}}
In a $2D$ spacetime, the Lorentz group is the pseudo-orthogonal group $O(1,1)$, which can be decomposed into $SO^+(1,1)\times Z_2\times Z_2$. This 1+1 dimensional Minkowski space can be described using split-complex or hyperbolic numbers, $z=\sigma + j\tau$, where $j^2=+1$ and its conjugate is $z^*=\sigma - j\tau$. The group $O(1,1)$ consists of transformations that preserve the inner product $zz^*$. In particular, the subgroup $SO^+(1,1)$ consists of hyperbolic rotations defined by $\{e^{j\theta} = \cosh(\theta)+j\sinh(\theta)\}$ where $\theta$ is real. The discrete Klein four-group $Z_2\times Z_2$ is represented by $\{z\longrightarrow \pm z$ and $z\longrightarrow \pm z^*\}$.

The compactification of the Lorentz group $O(1,1)$ for local spaces becomes $U(1)\times Z_2$. By replacing split-complex numbers with complex numbers, i.e., by replacing $j$ with the imaginary unit $i$, we obtain the circle group $U(1)=\{e^{i\theta}\}$. One of the generators of the Klein four-group ($z\longrightarrow -z$) is included in $U(1)$ as $\theta=\pi$, which is the reason why fermions in $2D$ are of Majorana type. The other generator (i.e., the complex conjugate $z\longrightarrow z^*$) is kept as mirror symmetry, which is the same as the chiral symmetry between holomorphic and anti-holomorphic modes in $2D$ models and string theory.

\subsection{$4D$ spacetime\label{app1.4}}
The Lorentz group for a $4D$ spacetime is $O(1,3)=SO^+(1,3)\times Z_2\times Z_2$. Its irreducible representations can be constructed from the irreducible representations $(m,n)$ of the group $SL(2,C)$, where $m,n=0,1/2,1,\ldots$. The two generators of its discrete subgroup $Z_2\times Z_2$ are often taken to be space inversion (parity) $P=diag(+1,-1,-1,-1)$ and time reversal $T=diag(-1,+1,+1,+1)$. Both of these transformations have determinant $-1$, resulting in a change of orientation.

Irreducible representations of $O(1,3)$ can be categorized into $(m,m)$ and $(m,n)\oplus(n,m)$ with $m\neq n$ \cite{ma2007}. Owing to the discrete symmetries, there are four inequivalent irreducible representations for $(m,m)$ and two inequivalent ones for $(m,n)\oplus(n,m)$. These can be combined to form the full representations of $O(1,3)$,
\begin{eqnarray}
&&(m,m)^+\oplus(m,m)^-\oplus(m,m)^{+\prime}\oplus(m,m)^{-\prime}, \\
&&(m,n)\oplus(n,m)\oplus(m,n)'\oplus(n,m)'.
\end{eqnarray}

When considering local DoFs in QFT, $O(1,3)$ must be compactified into $SU(2)\times SU(2)\times Z_2 \times Z_2$ where $SU(2)\times SU(2)$ inherits the same representations $(m,n)$ of the group $SL(2,C)$. The full representations when considering the discrete symmetries are the same as before. However, in QFT, more natural choices for generators of $Z_2 \times Z_2$ are $CP$ (particle\slash anti-particle symmetry) and $M$ (local orientation symmetry). $M$ is essential as no discrete operators in conventional QFT can change local orientation since all Dirac gamma matrices have determinant +1. Therefore, in general, elementary particles can be divided into four sectors: ordinary particles, ordinary anti-particles, mirror particles, and mirror anti-particles.

\section{Representations of elementary particle fields\label{app:rep}}

Starting in $2D$, supersymmetry becomes relevant and constrains the types of particle fields can appear in the Lagrangian due to the anti-commuting nature of fermionic coordinates in superspace. Only scalar, Majorana fermion, and gauge boson fields are allowed in $2D$. In $4D$ spacetime, Dirac fermion fields emerge replacing Majorana ones. Below, we will discuss two types of supermultiplets: gauge supermultiplets consisting of matter fermions and gauge bosons, and chiral supermultiplets consisting of fermions and scalars.

In $2D$ or higher dimensional spacetime, the base Lorentzian manifold has desired properties, such as almost-complex, pseudo-Riemannian, and symplectic structures. As a result, the base manifold is naturally associated with complex tangent\slash cotangent spaces. In particular, the exterior algebra of the complex (holomorphic) $n$-dimensional cotangent space $\mathbb{C}^n$ gives the fundamental representation of ordinary fermions in the following decomposition,
\begin{equation}
\wedge \mathbb{C}^n = \bigoplus_{i=0}^n \wedge^i \mathbb{C}^n
\end{equation}
and similarly, mirror fermions can be represented using the exterior algebra of the complex conjugate (anti-holomorphic) cotangent space $\bar{\mathbb{C}}^n$,
\begin{equation}
\wedge \bar{\mathbb{C}}^n = \bigoplus_{i=0}^n \wedge^i \bar{\mathbb{C}}^n.
\end{equation}

In $2D$ spacetime, with $n=1$, we have $\wedge^0 \mathbb{C}=\wedge^1 \mathbb{C}$. Therefore, the representation includes two fields of a holomorphic Majorana fermion $\lambda(z)$ and an anti-holomorphic Majorana fermion $\bar{\lambda}(\bar{z})$.

In the critical dimension of $D=10$, the base manifold is split into a $4D$ extended spacetime and a $6D$ curled-up CY space. Its complex (conjugate) cotangent spaces are split accordingly,
\begin{equation}
\mathbb{C}^5\longrightarrow \mathbb{C}^3\oplus\mathbb{C}^2, \;\;
\Bar{\mathbb{C}}^5\longrightarrow \Bar{\mathbb{C}}^3\oplus \mathbb{\Bar{C}}^2
\end{equation}
and the decomposition of the exterior algebra representation, for example, in the complex or ordinary sector, can then be written as,
\begin{equation}
\wedge (\mathbb{C}^3 \oplus \mathbb{C}^2) = \bigoplus_{i=0}^5 \bigoplus_{p+q=i} (\wedge^p \mathbb{C}^3 \otimes \wedge^q\mathbb{C}^2)
\end{equation}
which gives the representation of one generation fermions in the Standard Model as shown in Table \ref{tab:su5} (see details in Ref.~\cite{baez2005}).

Gauge bosons arise from the adjoint reps of the gauge groups that are simply the holonomy groups of the fiber and other curled-up spaces associated with the extended spacetime. Because of supersymmetry, these gauge bosons have the same number of DoFs as the fermions discussed above, i.e., $n_b=n_f$. Therefore, they can be grouped into the same gauge supermultiplet.

In $2D$ spacetime, the fiber space is simply the compactified string space. The requirement of complex structures naturally leads to a holonomy group of $U(1)$ that corresponds to two holomorphic and anti-holomorphic gauge bosons, $A(z)$ and $\Bar{A}(\bar{z})$. Together with the Majorana fermions $\lambda$ and $\bar{\lambda}$ discussed above, they form a gauge supermultiplet.

In the critical dimension of $D=10$, the underlying complex structures in all internal spaces require a universal $U_Y(1)$ gauge group for ordinary hypercharge and a $U'_Y(1)$ group for the mirror sector. CY mirror symmetry ensures the existence of two mirrored copies of the $SU(3)$ holonomy in the $6D$ CY space, providing the color gauge groups of $SU_c(3)$ for ordinary quarks and $SU'_c(3)$ for mirror quarks.

The 16 extra dimensions of the left-handed heterotic string must be compactified into a fiber space with an $SU_L(8)$ group that must be split with respect to the base manifold. In general, such a splitting for an $SU(N+M)$ group can be written as,
\begin{equation}\label{eq:split}
S(U(N)\times U(M)) = SU(N)\times SU(M)\times U(1) / Z_n \subset SU(N+M)
\end{equation}
where $n$ is the least common multiple of $N$ and $M$. In this case, we have $SU_L(8)\longrightarrow SU_L(2)\times SU_L(6)\times U_L(1)/Z_6$ where $SU_L(2)$ becomes the weak gauge group for the ordinary sector. The $U_L(1)$ group cannot be gauged due to its anomaly in $4D$ spacetime and ultimately breaks down to the global symmetry $U^{B-L}_V(1)$ for conservation of $B-L$. However,$SU_L(6)$, related to the quark space, can be further broken into an isospin group $SU_I(2)$ that can be gauged. Therefore, the complete gauge group for the ordinary sector is $U_Y(1)\times SU_L(2)\times SU_c(3)\times SU_I(2)$, which enables a gauge supermultiplet of $n_b=n_f=30$ involving one generation SM particles (excluding $\nu_R$ and $\bar{\nu}_R$) as shown in Table \ref{tab:su5}.

Alternatively, $SU_L(6)$ could be left as a completely global symmetry for quark flavors. Then the final gauge group is the well known SM group, $U_Y(1)\times SU_L(2)\times SU_c(3)$. Combining the DoFs from both gauge bosons and pNGBs from the flavor $SU_L(6)$ breaking, we have $n_b=90$ that leads to a pseudo-SUSY multiplet of $n_b=n_f=90$ for three generations of SM particles in the ordinary sector.

Similarly, the breakdown of the right-handed heterotic string's $SU_R(8)$ group gives us similar gauge groups in the mirror sector, i.e., $U'_Y(1)\times SU_R(2)\times SU'_c(3)\times SU'_I(2)$ in the UV limit that forms a gauge mirror supermultiplet of $n_b=n_f=30$ involving one generation of mirror fermions, and $U'_Y(1)\times SU_R(2)\times SU'_c(3)$ plus a global mirror flavor symmetry $SU_R(6)$ at low energies that gives a pseudo-SUSY multiplet of $n_b=n_f=90$ for three generations of mirror particles as shown in Table \ref{tab:rep}.

The gauge singlets of neutrinos and Higgs-like scalars from fermion condensation can form chiral supermultiplets. In $2D$, the Majorana fermions $\lambda$ and $\bar{\lambda}$, though in the gauge supermultiplet, carry no gauge charge and are therefore interaction-free. So they can also be part of the chiral supermultiplet. Their condensates produce a scalar field $\phi$ with two components of $\phi_L(z)$ and $\phi_R(\bar{z})$. This generates a chiral supermultiplet of $n_b=n_f=2$ as shown in Table \ref{tab:rep}.

Under the critical dimension of $D=10$, there are two possible choices of chiral supermultiplets. For the UV limit of one generation SM particles in the ordinary sector, there are two gauge singlets of $\nu_R$ and $\bar{\nu}_R$ that could be part of a possible chiral supermultiplet of $n_b=n_f=2$ with two scalar fields of $\phi_u$ and $\phi_d$. However, in reality, these condensates may not be formed and the corresponding scalar fields may remain massless due to constraints of supersymmetry.

The other choice is due to the emergence of three generations of particles and the breakdown of the global flavor $SU(6)$ group. At this low energy limit, six flavors of quarks can be condensed into six Higgs-like scalars $H_{u,d,c,s,t,b}$ in $SU_L(2)$ doublets that become real scalars $\phi_{u,d,c,s,t,b}$ after spontaneous symmetry breaking. Together with three families of neutrino singlets $\nu_R^{1,2,3}$ and $\bar{\nu}_R^{1,2,3}$, they form three chiral supermultiplets for the ordinary sector as shown in Table \ref{tab:rep} where similar supermultiplets for the mirror sector are also shown.

Based on these representations of supermultiplets, we can construct various supersymmetric mirror models under two different base manifold configurations with critical dimensions of $D=2$ and $D=10$, respectively.

\section{$2D$ Supersymmetric Mirror Models\label{app2}}

Using the $U(1)$ gauge supermultiplet discussed above, we write the following Lagrangian of the supersymmetric mirror model SMM2 under $2D$ spacetime $(t,\sigma)$,
\begin{equation}
\mathcal{L}_{\text{SMM2}} = -\frac{1}{4}F_{\alpha\beta}F^{\alpha\beta} + i\bar{\lambda}\rho^{\alpha}\partial_{\alpha}\lambda
\end{equation}
where the $U(1)$ gauge boson field $F_{\alpha\beta}=\partial_{\alpha}A_{\beta}-\partial_{\beta}A_{\alpha}$ and the Majorana-Weyl fermion field $\lambda$ are free, neutral, and massless. Obviously, this is simply a Lagrangian fixed in the conformal gauge of string theory on the $2D$ worldsheet.

Considering the $2D$ metric $h^{\alpha\beta}$ with a signature convention of $(-,+)$, we have $2D$ gamma matrices that obey the following anti-commutator,
\begin{equation}
\{\rho^{\alpha},\rho^{\beta}\} = -2h^{\alpha\beta}.
\end{equation}
In the real Majorana-Weyl representation, $\rho^{\alpha}$ are defined as,
\begin{equation}
\rho^{0}=
\begin{pmatrix}
0 & -i \\
i & 0
\end{pmatrix}
,\;\;\rho^1=
\begin{pmatrix}
0 & i \\
i & 0
\end{pmatrix}.
\end{equation}

In light-cone coordinates, $\sigma^{\pm}=t\pm\sigma$ and $\partial_{\pm} = \frac{1}{2}(\partial_t\pm\partial_{\sigma})$, the Lagrangian can be rewritten as,
\begin{equation}
\mathcal{L}_{\text{SMM2}} = -\frac{1}{4}(\partial_-A_+ - \partial_+A_-)^2 + i(\lambda_+\partial_-\lambda_+ + \lambda_-\partial_+\lambda_-)
\end{equation}
where $\lambda_+$ ($\lambda_-$) is a left-moving (right-moving) Majorana field.

After Wick rotation, the corresponding Euclidean action becomes,
\begin{equation}
\mathcal{S}^E_{\text{SMM2}} = \int d^2z\, \{\frac{1}{4}(\bar{\partial}A(z)-\partial \bar{A}(\bar{z}))^2+\lambda(z)\bar{\partial}\lambda(z) + \bar{\lambda}(\bar{z})\partial\bar{\lambda}(\bar{z})\}
\end{equation}
where $\partial=-i\partial_+$ and $\bar{\partial}=-i\partial_-$. The two holomorphic and anti-holomorphic modes of particles could be the origin of two distinct ordinary and mirror sectors in higher-dimensional spacetime.

Considering the real Majorana-Weyl representation and light-cone coordinates in $2D$ spacetime, the superspace can be denoted by $(\sigma^{\pm},\theta^{\pm})$, where $\theta^{\pm}$ are real fermionic coordinates.

The dual differential operators on superspace are,
\begin{eqnarray}
Q_{\pm} &=& -i\frac{\partial}{\partial \theta^{\pm}}+2\theta^{\pm}\partial_{\pm} \\
D_{\pm} &=& -i\frac{\partial}{\partial \theta^{\pm}}-2\theta^{\pm}\partial_{\pm} \nonumber
\end{eqnarray}
and they obey the following anti-commutation relations,
\begin{eqnarray}
\{Q_{\pm},Q_{\pm}\} &=& -4i\partial_{\pm},\;\; \{D_{\pm},D_{\pm}\} = 4i\partial_{\pm},\;\; \{Q,D\}=0, \\
\text{ and } \{Q_+,Q_-\}&=&0,\;\; \{D_+,D_-\}=0 \text{ for off-shell zero central charges. } \nonumber
\end{eqnarray}
A general superfield is defined as,
\begin{equation}
\Phi=\phi+i\theta^+\lambda_++i\theta^-\lambda_- + i\theta^+\theta^-F
\end{equation}
and a general action about the superfield can be written as,
\begin{eqnarray}
\mathcal{S} &=& \int d^2\sigma^{\pm} d\theta^- d\theta^+\, (\frac{1}{2}D_-\Phi D_+\Phi + V(\Phi)) \\
&=& \int d^2\sigma^{\pm} \, (2\partial_-\phi \partial_+\phi + i\lambda_+\partial_-\lambda_+ + i\lambda_-\partial_+\lambda_- -V''(\phi)\lambda_+\lambda_- +\frac{1}{2}(V'(\phi))^2) \nonumber
\end{eqnarray}
after eliminating the auxiliary field $F$. This gives a general formalism for the supersymmetric $N=(1,1)$ model SMM2b.

We can also rewrite the above in Euclidean space after Wick rotation to explicitly show the complex structure. The operators on superspace $(z,\bar{z},\theta,\bar{\theta})$ become,
\begin{eqnarray}
D_{\theta} &=& \frac{\partial}{\partial \theta} + \theta\partial_{z} \\
\bar{D}_{\bar{\theta}} &=& \frac{\partial}{\partial \bar{\theta}}+ \bar{\theta}\partial_{\bar{z}} \nonumber
\end{eqnarray}
and the superfield is rewritten as,
\begin{equation}
\Phi = \phi +\theta \lambda+\bar{\theta}\bar{\lambda} + \theta\bar{\theta}F.
\end{equation}
The SMM2b action then becomes,
\begin{eqnarray}
\mathcal{S}^E_{\text{SMM2b}} &=& \int d^2z d\theta d\bar{\theta}\, (D\Phi \bar{D}\Phi + V(\Phi)) \\
&=& \int d^2z \, \{\partial\phi \bar{\partial}\phi - \lambda\bar{\partial}\lambda - \bar{\lambda}\partial\bar{\lambda} -V''(\phi)\lambda\bar{\lambda} -\frac{1}{4}(V'(\phi))^2\}. \nonumber 
\end{eqnarray}
Note that $\phi = \phi_L(z) + \phi_R(\bar{z})$ is the sum of both holomorphic and anti-holomorphic states (i.e., ordinary and mirror scalars). The potential term could take an interesting Liouville form of $V(\phi) \sim e^{b\phi}$, which may be used to study the dynamics of cosmic inflation and black hole collapses.

\bibliography{mirror}

\end{document}